\documentclass[prd,aps,twocolumn,amsmath,amssymb,nofootinbib,preprintnumbers]
{revtex4}

\usepackage{graphicx}
\usepackage{dcolumn}
\usepackage{bm}
\usepackage{amsmath}
\usepackage{amsthm}
\usepackage{amsfonts}
\usepackage{euscript,bbm}
\usepackage{ifthen}
\usepackage{psfrag}
\usepackage{slashed}

\def\ls{\mathrel{\lower4pt\vbox{\lineskip=0pt\baselineskip=0pt
           \hbox{$<$}\hbox{$\sim$}}}}
\def\gs{\mathrel{\lower4pt\vbox{\lineskip=0pt\baselineskip=0pt
           \hbox{$>$}\hbox{$\sim$}}}}
\def\drawbox#1#2{\hrule height#2pt
\hbox{\vrule width#2pt height#1pt \kern#1pt
              \vrule width#2pt}
              \hrule height#2pt}

\def\Asym#1#2{\vcenter{\vbox{\drawbox{#1}{#2}
              \kern-#2pt       % line up boxes
              \drawbox{#1}{#2}}}}

\newcommand{\be}{\begin{equation}}
\newcommand{\ee}{\end{equation}}
\newcommand{\bea}{\begin{eqnarray}}
\newcommand{\eea}{\end{eqnarray}}

\newcommand{\gsim}{\lower.7ex\hbox{$\;\stackrel{\textstyle>}{\sim}\;$}}
\newcommand{\lsim}{\lower.7ex\hbox{$\;\stackrel{\textstyle<}{\sim}\;$}}

\newcommand{\vo}{\mathcal{V}}

\newcommand{\ben}{\begin{enumerate}}
\newcommand{\een}{\end{enumerate}}
\newcommand{\bei}{\begin{itemize}}
\newcommand{\eei}{\end{itemize}}
\newcommand{\mc}{\mathcal}

\begin{document}

\title{Non-thermal Dark Matter in String Compactifications}

\author{Rouzbeh Allahverdi$^{1}$}
\author{Michele Cicoli$^{2,3,4}$}
\author{Bhaskar Dutta$^{5}$}
\author{Kuver Sinha$^{6}$}

\affiliation{$^{1}$~Department of Physics and Astronomy, University of New Mexico, Albuquerque, NM 87131, USA \\
$^{2}$~Dipartimento di Fisica ed Astronomia, Universit\`a di Bologna, via Irnerio 46, 40126 Bologna, Italy \\
$^{3}$~INFN, Sezione di Bologna, 40126 Bologna, Italy \\
$^{4}$~Abdus Salam ICTP, Strada Costiera 11, Trieste 34014, Italy \\
$^{5}$~Department of Physics and Astronomy, Texas A\&M University, College Station, TX 77843-4242, USA \\
$^{6}$~Department of Physics, Syracuse University, Syracuse, NY 13244, USA
}

\begin{abstract}
Non-thermal cosmological histories are capable of greatly increasing the available parameter space of different particle physics dark matter (DM)
models and are well-motivated by the ubiquity of late-decaying gravitationally coupled scalars in UV theories like string theory.
A non-thermal DM model is presented in the context of LARGE Volume Scenarios in type IIB string theory.
The model is capable of addressing both the moduli-induced gravitino problem as well as
the problem of overproduction of axionic dark radiation and/or DM. We show that the right abundance of neutralino DM can be obtained in both thermal under and overproduction cases for DM masses between ${\cal O}({\rm GeV})$ to ${\cal O}({\rm TeV})$. In the latter case the contribution of the QCD axion to the relic density is totally negligible, while in the former case it can be comparable to that of the neutralino thus resulting in a multi-component DM scenario.
\end{abstract}
MIFPA-13-23$\qquad$CETUP2013-003
\maketitle

\section{Introduction}

The standard paradigm of thermal dark matter (DM) assumes DM in thermal equilibrium following an initial inflationary era.
Subsequently, the DM particle drops out of thermal equilibrium and its abundance freezes out when annihilation becomes inefficient at a temperature of order $T_{\rm f} \simeq m_{\rm DM}/20$. Due to the lack of direct observations of the history of the universe before Big Bang Nucleosynthesis (BBN),
it is important to go beyond the standard thermal paradigm.
In fact, non-thermal DM is well-motivated both from a bottom-up and a top-down point of view.

From a bottom-up approach, non-thermal DM scenarios vastly enlarge the parameter space available in particle physics models.
The most obvious example is the Minimal Supersymmetric Standard Model (MSSM), where neutralino DM candidates typically give too much
(Bino DM) or too little (Higgsino/Wino DM) relic density. The correct thermal relic density is only satisfied for very specific cases: $(i)$ Bino DM with co-annihilation, Higgs resonance, or $t$-channel slepton exchange or $(ii)$ Well-tempered Bino/Higgsino or Bino/Wino DM. The first case is typically fine-tuned, while the second is being increasingly constrained by direct detection experiments. There is also the possibility of multi-component Higgsino or Wino DM, but this requires additional physics. In a non-thermal scenario~\cite{MR}, both cases with thermal under (Wino/Higgsino) and overabundance (Bino) can be accommodated. %\cite{GelminiGondolo}. 
Non-thermal production of Wino DM \cite{MoroiRandall} provides an explicit example. Another important example is pure Higgsino DM \cite{Allahverdi:2012wb}, which is motivated by naturalness conditions \cite{Natural}.

Furthermore, light DM with mass $\sim \mathcal{O}(10)$ GeV, motivated by results from recent direct detection experiments \cite{Agnese:2013dwa, cogent,cresst,dama}, typically has an annihilation cross-section that is smaller in the context of most models due to the exchange of O(TeV)  particles \cite{Allahverdi:2013tca} which leads to overabundance of dark matter in the current epoch in the thermal scenarios. Also, for DM with mass $\lsim \, 40$ GeV the annihilation cross-section is  constrained to be less than the thermally required value  by the gamma ray flux from dwarf spheroidal galaxies and galactic center \cite{fermi,Center}.

From a top-down approach, the ubiquity of gravitationally coupled moduli in string theory makes the scenario of a late-decaying scalar quite generic.
Late-time decay will typically erase any previously produced DM relic density as well as baryon asymmetry, necessitating non-thermal production. The modulus should decay before BBN and late enough to produce interesting effects on IR physics. Hence non-thermal physics requires
$T_{\rm BBN} \lesssim T_{\rm rh} < T_{\rm f}$, where $T_{\rm rh}$ is the modulus reheat temperature and $T_{\rm BBN} \simeq 3$ MeV is the lower bound required by the success of BBN.
This typically places upper and lower bounds on the mass of the modulus. However, given that the scale of soft masses also depends on the moduli masses,
the requirement of TeV-scale supersymmetry (SUSY) to solve the hierarchy problem, generically forces the moduli to be light enough to decay at
temperatures below $T_{\rm f}$. Thus, from the point of view of string theory, non-thermal DM scenarios seem to be more
generic than standard thermal ones \cite{Bobby}.

The purpose of this paper is to explore non-thermal DM in string compactifications, specifically sequestered models
in the context of type IIB LARGE Volume Scenarios (LVS) \cite{LVS}. Several problems associated with non-thermal scenarios are readily addressed in this context: the moduli-induced gravitino problem \cite{gravProbl} and the overproduction of axionic dark radiation (DR) \cite{DR1,DR2} and DM \cite{AxionDM}.
Broadly speaking, the gravitino problem is solved by kinematic suppression since in these models
the gravitino is much heavier than the modulus while still yielding TeV-scale superpartners.
Axionic DR overproduction can be addressed either by suitable coupling of the modulus to the visible sector,
or by removing dangerous axions via anomalous $U(1)$ symmetries. On the other hand, axionic DM overproduction
can be avoided either by a sufficiently low reheat temperature, or by considering open string axions
with a low decay constant $f_a\simeq 10^{12}$ GeV. Moreover, it is possible to accommodate cases
of both thermal DM over and underproduction. Both cases can be realized using neutralinos for masses between ${\cal O}$(GeV) to ${\cal O}$(TeV).  In the underproduction case, if needed, the QCD axion can be utilized to satisfy the abundance.

The plan of the paper is as follows. In Section \ref{challenges} we describe non-thermal scenarios from moduli decays
and their main challenges. In Section \ref{LVSsection} we briefly review sequestered LVS models where these problems are addressed.
Finally in Section \ref{DMvolmodulus} we work out the production of non-thermal DM in these models before ending with our Conclusions.

\section{Challenges for Non-thermal Scenarios}
\label{challenges}

There are two ubiquitous problems of any string compactification which are particularly relevant for us:
\ben
\item \emph{Cosmological moduli problem (CMP)}\cite{CMP}: the moduli start oscillating around their minimum
at $H_{\rm osc}\sim m_{\rm mod}$ with an initial displacement at the end of inflation of order $M_{\rm P}$.
Redshifting as matter, they quickly come to dominate the energy density,
and so they reheat our universe when they decay. Being gravitationally coupled scalars,
they decay very late when $H_{\rm dec}\sim \epsilon^2 m_{\rm mod}$ where $\epsilon \sim m_{\rm mod}/M_{\rm P}$.
In order to prevent any distortion of the successful BBN predictions, the resulting reheat temperature
$T_{\rm rh}\sim \epsilon^{1/2} m_{\rm mod}$ must be above $T_{\rm BBN}$.
This sets a lower bound on the moduli masses of order $m_{\rm mod}\gtrsim 10$ TeV.

\item \emph{Axionic DM overproduction}\cite{AxionDM}: string compactifications give rise in 4D to many axion-like
particles whose number is controlled by the topology of the extra-dimensions, and it is generically of order a few hundred.
Some of these axions can be projected out from the low-energy spectrum, can be eaten up by anomalous $U(1)$s or
can get large masses from non-perturbative effects, but generically some of them will remain light \cite{LVSaxions}.
One of these light axions, if it has the right coupling to QCD, can then play the r\^ole of the QCD axion. Its decay constant
turns out to be set by the string scale $f_a\simeq M_{\rm s}/\sqrt{4\pi}$ which is generically very high, $M_{\rm s}\gtrsim 10^{15}$ GeV,
resulting in the overproduction of axionic cold DM for $f_a\gtrsim 10^{14}$ GeV.
\een

There seems to be a tension between these two problems since in order to make the moduli heavier than $10$ TeV, one has
to raise all the scales in the model, so increasing also the axion decay constant.
However, if $m_{\rm mod}> 10$ TeV, the heavy moduli decaying in the window $T_{\rm BBN} \lesssim T_{\rm rh} < T_{\rm f}$
would dilute any previous relic. In particular:
\ben
\item Axionic DM is diluted if $T_{\rm rh}<\Lambda_{\rm QCD}\simeq 200$ MeV, so avoiding any overproduction \cite{Fox}.
The maximum dilution is obtained for $T_{\rm rh}$ very close to $T_{\rm BBN}$ allowing a decay constant
of order $f_a\simeq 10^{14}$ GeV without fine-tuning the initial misalignment angle.

\item Standard thermal DM is also diluted if $T_{\rm rh}<T_{\rm f}\simeq \mc{O}(10)$ GeV. DM would then
be produced non-thermally by the moduli decay. There are two viable mechanisms
depending on the value of $T_{\rm rh}$. If $T_{\rm rh}$ is close to $T_{\rm f}$,
one has to consider the `annihilation scenario', with a very abundant initial production
of Wino- or Higgsino-like DM particles and a subsequent very efficient annihilation process.
On the other hand, if $T_{\rm rh}$ is close to $T_{\rm BBN}$, a smaller initial abundance of DM particles is produced
and strong annihilation effects would lower it even further. Thus, in this case, one is in the `branching scenario'
where the right amount of Bino-like DM is produced directly from the moduli decay without any subsequent annihilation.

\item Any matter-antimatter asymmetry produced before the moduli decay would also be diluted.
This could be a welcomed effect if baryogenesis occurs in the early universe via the Affleck-Dine
mechanism which tends to be too efficient. On the other hand, if no matter-antimatter asymmetry
is left over after the moduli decay, this scenario opens up the possibility to
explain the cosmic coincidence puzzle, i.e. why the DM abundance is of the same order of the one of ordinary baryons.
In fact, both of them could be produced from the moduli decay into
new heavy coulored particles with baryon number and CP-violating couplings \cite{Clado}.
\een
In summary, the moduli decay can solve the problem of axionic DM overproduction,
can give rise to non-thermal DM, and finally can be responsible also for baryogenesis.
This seems to be the generic situation for string compactifications where the moduli are stabilised
by breaking SUSY and no CMP is present,
since in this case both the mass of the lightest modulus $m_{\rm light}$ and
the scale of the soft terms $M_{\rm soft}$ generated by gravity mediation are controlled by the gravitino mass:
$M_{\rm soft}\lesssim m_{\rm light}\sim m_{3/2}$. Thus the reheat temperature
from the lightest modulus decay scales as $T_{\rm rh}\gtrsim M_{\rm soft}\sqrt{M_{\rm soft}/M_{\rm P}}$.
In order to make it larger than $\Lambda_{\rm QCD}$,
the soft terms should be larger than $\mc{O}(500)$ TeV, so losing the possibility
to rely on low-energy SUSY to solve the hierarchy problem.
A possibly different situation is the one where the visible sector is `sequestered'
from SUSY breaking so that $M_{\rm soft}\ll m_{\rm light}$
since in this case TeV-scale SUSY could be compatible
with a reheat temperature above $T_{\rm f}$.
In what follows we shall however show that this is not the case.

Let us finally mention that the decay of the moduli could in principle also introduce two problems:
\bei
\item \emph{Gravitino problem}\cite{gravProbl}: If $m_{3/2}< m_{\rm light}$
the gravitino is produced by the light modulus decay. If $m_{3/2}\lesssim 10$ TeV,
the gravitino decays after BBN, otherwise if $m_{3/2}\gtrsim 10$ TeV,
the gravitini could annihilate into DM causing its overproduction.

\item \emph{Dark radiation overproduction}\cite{DR1,DR2}: The moduli are gauge singlets and so
they do not prefer to decay into visible sector fields. Thus,
if light hidden sector degrees of freedom like axion-like particles exist,
the branching ratio into them could not be negligible, so giving a number of
effective relativistic species which is above the tight bounds from cosmological observations,
$\Delta N_{\rm eff}\simeq 0.5$ \cite{Neff}.
\eei
In the next section we present a string model where the moduli masses
and couplings can be computed explicitly and these two problems can
be easily evaded.

\section{Sequestered LVS Models}
\label{LVSsection}

A very promising moduli stabilisation mechanism in type IIB string theory is
the LARGE Volume Scenario \cite{LVS}. In this framework,
all the moduli are fixed by background fluxes, D-terms from anomalous $U(1)$s,
and the interplay of non-perturbative and $\alpha'$ effects.
The simplest realisation involves an internal volume of the form
(for explicit constructions see \cite{ExplicitLVS}):
\be
\vo= \tau_{\rm big}^{3/2}-\tau_{\rm np}^{3/2}-\tau_{\rm inf}^{3/2}-\tau_{\rm vs}^{3/2}\,,
\ee
where the $\tau$'s are K\"ahler moduli parameterising the size of internal 4-cycles.
The visible sector (a chiral MSSM- or GUT-like theory)
is built via space-time filling D3-branes sitting at the singularity
obtained by shrinking $\tau_{\rm vs}$ to zero size by D-terms \cite{seqLVS}.
On the other hand, the cycle $\tau_{\rm np}$ supports non-perturbative
effects which fix it in terms of the string coupling: $\langle\tau_{\rm np}\rangle \simeq g_s^{-1}$.
For $g_s\simeq 0.1$ in the perturbative regime, $\tau_{\rm np}$ is of order $10$ in string units.
The `big' cycle $\tau_{\rm big}$ is instead stabilised by $\alpha'$ plus non-perturbative effects
at $\langle\vo\rangle \simeq \langle \tau_{\rm big}\rangle^{3/2} \simeq W_0 \, e^{2\pi/(N g_s)}$
where $W_0$ is the flux-generated superpotential and $N$ is the rank of the condensing
gauge group. This minimum breaks SUSY spontaneously. Minkowski vacua can be obtained
by either D-terms \cite{Dup} or non-perturbative effects at singularities \cite{NPup}.
The modulus $\tau_{\rm inf}$ behaves similarly to $\tau_{\rm np}$,
and by displacing it from its minimum, it can drive 60 e-folds of inflation,
generating a red spectrum and the right amount of density perturbations for $\vo \simeq 10^7$ \cite{KMI}.

Since the volume is exponentially large,
it is easy to generate such large numbers for natural values of the underlying parameters.
An important scale in the model is the mass of the soft terms $M_{\rm soft}$ generated by gravity mediation.
Given that $\langle\tau_{\rm vs}\rangle=0$, this modulus has a vanishing F-term as opposed
to all the other moduli which develop non-zero F-terms. As a consequence, the visible sector
is sequestered and the soft terms are significantly suppressed with respect to $m_{3/2}$.
All the relevant energy scales in the model are set by value of $\vo$ \cite{seqLVS}:
\bei
\item Reduced Planck scale: $M_{\rm P} = 2.4 \times 10^{18}$ GeV
\item GUT-scale: $M_{\rm GUT} \simeq {M_{\rm P}}/{\vo^{1/3}}$,
\item String-scale and $\tau_{\rm vs}$: $M_{\rm s} \simeq m_{\tau_{\rm vs}} \simeq {M_{\rm P}}/{\vo^{1/2}}$,
\item Kaluza-Klein scale: $M_{\rm KK} \simeq {M_{\rm P}}/{\vo^{2/3}}$,
\item Inflaton and $\tau_{\rm np}$: $m_{\tau_{\rm inf}} \simeq m_{\tau_{\rm np}} \simeq m_{3/2} \ln \vo$
\item Gravitino mass: $m_{3/2} \simeq W_0\,{M_{\rm P}}/{\vo}$,
\item Big modulus: $m_{\tau_{\rm big}} \simeq {m_{3/2}}/{\vo^{1/2}}$,
\item Soft-terms: $M_{\rm soft} \simeq m_{3/2}/\vo$.
\eei
Setting $W_0 \sim 0.1$ and $\vo \simeq 10^7$, one obtains $M_{\rm GUT} \simeq 10^{16}$ GeV,
$M_{\rm s} \simeq 10^{15}$ GeV, $M_{\rm KK} \simeq 5 \times 10^{13}$ GeV,
$m_{\tau_{\rm inf}} \simeq m_{\tau_{\rm np}} \simeq 10^{11}$ GeV,
$m_{3/2}\sim 10^{10}$ GeV, $m_{\tau_{\rm big}} \simeq 5 \times 10^6$ GeV
and $M_{\rm soft} \simeq 1$ TeV.

An interesting observation is that for $\vo \simeq 10^7$,
one can get both inflation and low-energy SUSY.
Moreover, all the moduli are heavy, and so
there is no CMP. The gravitino problem is also
avoided since $m_{3/2} \gg m_{\tau_{\rm big}}$.

As far as the moduli couplings are concerned,
the leading decay channels for $\tau_{\rm big}$ are to Higgses and
closed string axions. Denoting as $\phi$ the canonically normalised modulus $\tau_{\rm big}$,
the various decay rates of this modulus are (see \cite{DR1} for details):
\bei
\item \emph{Decays to Higgs bosons}: the decays $\phi \to H_u H_d$ are induced by the Giudice-Masiero term in the K\"ahler potential,
$K \supset Z\,\frac{H_u H_d}{2\tau_{\rm big}}$, where $Z$ is an $\mc{O}(1)$ parameter. The corresponding decay rate is:
\be
\label{Z}
\Gamma_{\phi \to H_u H_d} = \frac{2 Z^2}{48 \pi} \frac{m_{\phi}^3}{M_{\rm P}^2}\,.
\ee

\item \emph{Decays to bulk axions}: the axionic partner $a_{\rm big}$ of the big modulus is almost massless, and
so $\tau_{\rm big}$ can decay into this particle with decay width:
\be
\Gamma_{\phi \to a_{\rm big} a_{\rm big}} = \frac{1}{48 \pi} \frac{m_{\phi}^3}{M_{\rm P}^2}\,.
\ee

\item \emph{Decays to local closed string axions}: $\tau_{\rm big}$ can decay also to
closed string axions $a_{\rm loc}$ localised at the singularity hosting the visible sector with decay rate:
\be
\Gamma_{\phi \to a_{\rm loc} a_{\rm loc}} = \frac{9}{16} \frac{1}{48 \pi} \frac{m_{\phi}^3}{M_{\rm P}^2}\,.
\ee

\item \emph{Decays to gauge bosons}: given that the holomorphic gauge kinetic function does not depend on
$\tau_{\rm big}$ due to the localisation of the visible sector at a singularity, this modulus
couples to gauge bosons only due to radiative corrections, inducing a loop suppressed decay width:
\be
\Gamma_{\phi \to A^{\mu} A^{\mu}}  =\lambda \left(\frac{\alpha_{\rm SM}}{4 \pi} \right)^2 \frac{m_{\phi}^3}{M_{\rm P}^2}\,,
\label{phigauge}
\ee
where $\lambda\sim \mc{O}(1)$ and $\alpha_{\rm SM}$ is the corresponding coupling constant.
\item \emph{Decays to other visible sector fields}: the decays to matter scalars, fermions, gauginos and Higgsinos
(commonly denoted as $\psi$) are all mass suppressed since their corresponding decay rates scale as:
\be
\Gamma_{\phi \to \psi \psi} \simeq \frac{M_{\rm soft}^2 m_\phi}{M_{\rm P}^2} \ll \frac{m_{\phi}^3}{M_{\rm P}^2}\,.
\ee

\item \emph{Decays to local open string axions}: the decays to light open string axions which
are the phase $\theta$ of a matter field $C=\rho \,e^{i\theta}$ arise from the coupling:
\be
\left(\frac{\langle \rho \rangle}{M_{\rm P}}\right)^2 \phi \theta \Box \theta\,,
\ee
and so are suppressed by both the tiny mass of the axion and the fact that $(\langle\rho\rangle/M_{\rm P})^2 \simeq 1/\vo \ll 1$.
\eei

As pointed out in \cite{DR1}, the unsuppressed decays to bulk and local closed string axions
can cause problems with DR overproduction. However, globally consistent brane constructions
in explicit Calabi-Yau examples have revealed that both the light bulk axion $a_{\rm big}$ and
all the local closed string axions $a_{\rm loc}$ tend to be eaten up by anomalous $U(1)$s \cite{ExplicitLVS}.
We shall therefore not consider it as a
serious problem. On the other hand, the QCD axion can have different phenomenological features
according to its origin as a closed or an open string mode \cite{LVSaxions}:
\bei
\item \emph{Closed string QCD axion}: If at least one local closed string axion is not eaten
up by any anomalous $U(1)$, it can play the r\^ole of the QCD axion.
Given that its decay constant is set by the string scale, $f_a \simeq M_{\rm s}/\sqrt{4\pi}\simeq 10^{14}$ GeV,
it needs to be diluted by the decay of $\tau_{\rm big}$ (otherwise one has to fine-tune
the initial misalignment angle). Moreover, one has to make sure that
it does not cause any problem with DR overproduction.

\item \emph{Open string QCD axion}: If the QCD axion is the phase of a matter field,
then the modulus decay rate to this particle is subleading, so leading to no DR production.
Furthermore, in this case the axion decay constant gets reduced with respect to the string scale,
$f_a \simeq M_{\rm s}/\vo^{\alpha}$ with $0< \alpha < 1$. For $\alpha=1/2$, one has $f_a\simeq 10^{11}$ GeV,
perfectly within the QCD axion allowed window $10^9 {\rm GeV}\lesssim f_a \lesssim 10^{12} {\rm GeV}$.
\eei

\section{Non-thermal Dark Matter from Lightest Modulus Decay}
\label{DMvolmodulus}

The lightest modulus $\tau_{\rm big}$ serves as the source of non-thermal DM.
The modulus interacts gravitationally with other fields, leading to a decay width given by:
\be
\Gamma_\phi = \frac{c}{2\pi} \frac{m_\phi^3}{M_{\rm P}^2}\,,
\ee
where $c$ is a constant that depends on the decay modes of the modulus.
The modulus decays when $H \sim \Gamma_\phi$ and reheats the universe to a temperature:
\be
T_{\rm rh} = c^{1/2} \left(\frac{10.75}{g_*}\right)^{1/4} \left( \frac{m_\phi}{50\, {\rm TeV}}\right)^{3/2}\, T_{\rm BBN} \,, 
\nonumber \label{Tr}
\ee
where $T_{\rm BBN} \simeq 3 ~ {\rm MeV}$ and $g_*$ is the number of relativistic degrees of freedom at $T_{\rm rh}$.
The modulus decay dilutes the abundance of existing DM particles by at least a factor of order $(T_{\rm f}/T_{\rm rh})^3$,
where $T_{\rm f}$ is freeze-out temperature of DM annihilation.
This can be easily a factor of $10^6$ or larger, hence requiring DM production from modulus decay
in order to explain the DM content of the universe.\footnote{For a scenario with a mild dilution of thermally overproduced DM by modulus decay, see \cite{Hooper}.} The abundance of DM particles produced in this way is:
\begin{eqnarray}
\label{dmdens}
{n_{\rm DM} \over s} = {\rm min} \left[\left({n_{\rm DM} \over s}\right)_{\rm obs}
{\langle \sigma_{\rm ann} v \rangle_{\rm f}^{\rm th} \over \langle \sigma_{\rm ann} v \rangle_{\rm f}} \left({T_{\rm f} \over T_{\rm rh}}\right),
Y_{\phi}~ {\rm Br}_{\rm DM} \right] \, , \nonumber \\
\,
\end{eqnarray}
where $\langle \sigma_{\rm ann} v \rangle_{\rm f}^{\rm th}\simeq 3 \times 10^{-26} {\rm cm}^3\,{\rm s}^{-1}$
is the value needed in the thermal case to match the observed DM abundance:
\be
\label{obs}
\left({n_{\rm DM} \over s}\right)_{\rm obs} \simeq 5 \times 10^{-10} ~ \left({1 ~ {\rm GeV} \over m_{\rm DM}}\right),
\ee
whereas the yield of particle abundance form $\phi$ decay is:
\be
Y_\phi \equiv {3 T_{\rm rh} \over 4 m_\phi} = {0.9 \over \pi} ~ \sqrt{\frac{c\, m_\phi}{M_{\rm P}}}\,.
\label{yield}
\ee
${\rm Br}_{\rm DM}$ denotes the branching ratio for $\phi$ decays into $R$-parity odd particles which subsequently decay to DM.

The first term on the right-hand side of eq.~(\ref{dmdens}) is the \emph{Annihilation Scenario}
since DM particles produced from the modulus decay undergo some annihilation.
This can happen when $\langle \sigma_{\rm ann} v \rangle_{\rm f} = \langle \sigma_{\rm ann} v \rangle_{\rm f}^{\rm th}\, (T_{\rm f}/T_{\rm rh})$.
Since $T_{\rm rh} < T_{\rm f}$, this scenario can yield the correct DM abundance only if
$\langle \sigma_{\rm ann} v \rangle_{\rm f} > \langle \sigma_{\rm ann} v \rangle_{\rm f}^{\rm th}$ (as for Higgsino DM).

The second term on the right-hand side of eq.~(\ref{dmdens}) is the \emph{Branching Scenario}
where the residual annihilation of DM particles is inefficient and the final DM abundance
is the same as that produced form the modulus decay. This happens if
$\langle \sigma_{\rm ann} v \rangle_{\rm f} < \langle \sigma_{\rm ann} v \rangle_{\rm f}^{\rm th}
\, (T_{\rm f}/T_{\rm rh})$. We note that this is always the case for
$\langle \sigma_{\rm ann} v \rangle_{\rm f} < \langle \sigma_{\rm ann} v \rangle_{\rm f}^{\rm th}$ (like in the case of Bino DM).
It may also happen for $\langle \sigma_{\rm ann} v \rangle_{\rm f} > \langle \sigma_{\rm ann} v \rangle_{\rm f}^{\rm th}$
if $T_{\rm rh}/T_{\rm f}$ is too small.

The Fermi results, based on data from dwarf spheroidal galaxies and the galactic center,
have already placed tight constraints on the ``annihilation scenario''.
The limits from dwarf galaxies \cite{fermi} indicate that $T_{\rm f} \lesssim 30 \,T_{\rm rh}$
for $m_{\rm DM} > 40$ GeV, which implies $T_{\rm rh} > 70$ MeV. For $m_{\rm DM} < 40$ GeV,
the Fermi bounds require $\langle \sigma_{\rm ann} v \rangle_{\rm f} < \langle \sigma_{\rm ann} v \rangle_{\rm f}^{\rm th}$,
if DM annihilates into $b {\bar b}$ with S-wave domination, implying that the ``annihilation scenario'' cannot work in this case.
The constraints become stronger when galactic center data~\cite{Center} are included.\footnote{See also \cite{NonThCosm} for the effect of a non-thermal phase on inflationary observables.}

As a result, the ``branching scenario'' is strongly preferred as the only option in the mass range $m_{\rm DM} < 40$ GeV.
Since $5 \times 10^{-3} \lesssim {\rm Br}_{\rm DM} \lesssim 1$, with the lower bound set by three-body decay of
$\phi$ into $R$-parity odd particles~\cite{Clado}, we need $Y_\phi \lesssim 10^{-8}$ in order to obtain the correct DM abundance within this scenario.
For $m_\phi \simeq 5\times 10^6$ GeV, $Y_\phi \lesssim 10^{-8}$ requires $T_{\rm BBN}\lesssim T_{\rm rh} \lesssim 70$ MeV.

Based on the above arguments, we find that there are two interesting regimes for $T_{\rm rh}$:
\ben
\item Annihilation scenario for $T_{\rm f}/30 \lesssim T_{\rm rh} < T_{\rm f}$;
\item Branching scenario for $T_{\rm BBN} \lesssim T_{\rm rh} \lesssim 70$ MeV.
\een
We shall now discuss these two cases in more detail.

\subsection{Annihilation Scenario for High $T_{\rm rh}$}

In the regime $T_{\rm f}/30 \lesssim T_{\rm rh} < T_{\rm f}$ the annihilation scenario is at work.
As we have seen in section \ref{LVSsection}, $\phi$ decays primarily to Higgses, giving $c=Z^2/12$.
Inserting this value and in eq. (\ref{Tr}) together with the modulus mass $m_\phi \simeq 5\times10^6$ GeV
that gives TeV-scale SUSY, we find a reheat temperature of order $T_{\rm rh}\simeq 0.8\,Z$ GeV.

Focusing on situations where bulk axions are removed from the spectrum, the QCD axion can either be a closed
or an open string mode:
\ben
\item The QCD axion is a local closed string mode $a_{\rm loc}$ with $f_a \sim 10^{14}$ GeV:
\bei
\item $\phi\to a_{\rm loc} a_{\rm loc}$ is a leading decay channel, and so we need
to suppress the contribution to $\Delta N_{\rm eff} \simeq 1/Z^2$ \cite{DR1}.
In order to have $\Delta N_{\rm eff}\simeq 0.5$ we need $Z\simeq \sqrt{2}$,
which gives $T_{\rm rh}\simeq 1$ GeV.

\item In order to have $T_{\rm rh}<T_{\rm f}$, one needs $m_{\rm DM}> 20\, T_{\rm rh}\simeq 20$ GeV.

\item The reheat temperature is larger than the QCD scale, $T_{\rm rh}\simeq 1\,{\rm GeV} >\Lambda_{\rm QCD}$, and
so axion cold DM is not diluted. Hence one has either to tune the initial misalignment angle or to
remove $a_{\rm loc}$ from the spectrum with the help of an anomalous $U(1)$ (the QCD axion has then to be an open string mode).

\item Tuning the misalignment angle suitably it is possible to make
 multicomponent DM (Wino/Higgsino-like + closed string axions) \footnote{Considering different astrophysical observations, the viability of non-thermal Wino DM may be very constrained \cite{winodmnonth}}.
\eei

\item The QCD axion is an open string mode $\theta$ with $f_a\simeq 10^{11}$ GeV.
\bei
\item $\phi\to \theta\theta$ is a subleading decay channel, and so no DR is produced.

\item Due to the high value of $T_{\rm rh}$
the modulus decay does not result in any dilution of axion oscillations,
but since $f_a$ is intermediate, we do not need to tune the initial misalignment angle
to avoid axionic DM overproduction.

\item Again DM is generically multi-component (Wino/Higgsino-like + open string axions). The open string axion contribution 
to the DM abundance reduces as $f_a$ becomes smaller than $10^{12}$ GeV.
\eei
\een

In summary, in the annihilation scenario the lightest neutralino (Wino/Higgsino type) can satisfy the relic density. If, however, the abundance is small, the QCD axion can be utilized to form a multi-component DM scenario \cite{Howie}.

\subsection{Branching Scenario for Low $T_{\rm rh}$}

In order to have a low $T_{\rm rh}$ ($3\,{\rm MeV} \lesssim T_{\rm rh} \lesssim 70$ MeV)
the modulus decay width has to be very small. However, as we have seen in the previous section,
if the QCD axion is a closed string mode,
then we will need $Z \geq \sqrt{2}$ in order to avoid the DR problem.
This in turn sets $T_{\rm rh}\gtrsim 1$ GeV. In order to lower $T_{\rm rh}$
one could consider smaller values of $m_\phi$ which would however imply $M_{\rm soft} \ll 1$ TeV.
Hence the only way-out is to focus on cases where the closed string axions are absorbed by anomalous $U(1)$s,
and the QCD axion is realised as an open string mode $\theta$. Due to its suppressed coupling to $\phi$,
$\theta$ does not cause any DR problem, allowing very low values of $T_{\rm rh}\ll \Lambda_{\rm QCD}$.
Thus in this case the modulus decay will dilute the axion oscillations, leading
to a negligible contribution of the QCD axion to DM.

There are two ways to lower the reheat temperature:
\ben
\item If the Giudice-Masiero term is forbidden by some symmetries then $Z=0$.
In this case the leading decay channel is to gauge bosons via a two-body final state
with a loop-suppressed decay rate, giving $c=\lambda \frac{\alpha_{\rm SM}^2}{8 \pi}$ (see eq.~(\ref{phigauge})).
If $\lambda \simeq 1$, $\alpha_{\rm SM}\simeq 1/137$, and $m_\phi \simeq 5\times10^6$ GeV,
the reheat temperature is $T_{\rm rh}\simeq 4$ MeV (slightly above BBN),
giving $Y_\phi \simeq 6 \times 10^{-10}$.
Two-body decays to gauginos and other MSSM particles are instead mass suppressed in this case.
However, gauginos are inevitably produced in three-body decays of the modulus
(e.g., $\phi \rightarrow 1 ~ {\rm gluon} + 2 ~ {\rm gluinos}$) with ${\rm Br}_{\rm DM} \sim 5 \times 10^{-3}$.
Then ${\rm Br}_{\rm DM} \simeq 5 \times 10^{-3}$ results in a DM abundance
which matches the observed value for $m_{\rm DM} \simeq 165$ GeV.
The DM mass is inversely proportional to $Y_\phi \propto \sqrt{\lambda m_\phi}$.
Larger values of $m_\phi$ would require smaller values of $m_{\rm DM}$ but
in this situation the soft terms would become larger than the TeV-scale.
On the other hand, smaller values of $m_\phi$ would imply larger values
of $m_{\rm DM}$ but then $M_{\rm soft}\ll 1$ TeV. Hence we shall keep $m_\phi$
fixed at $m_\phi\lesssim 5\times 10^6$ GeV and try to vary $\lambda$.
The requirement $T_{\rm rh}\gtrsim 3$ MeV
implies $\lambda \gtrsim 0.01$ and in turn $m_{\rm DM}\lesssim 900$ GeV.

\item In the absence of symmetries forbidding the decay of $\phi$ to Higgses,
it is still possible to have low $T_{\rm rh}$ for $Z\simeq 0.1$.
In this case, for $m_\phi \simeq 5\times10^6$ GeV, we would have $T_{\rm rh}\simeq 80$ MeV,
which implies $m_{\rm DM} \simeq 10$ GeV. Larger values of $m_{\rm DM}$ require
smaller values of $Z$ keeping $m_\phi$ fixed to get TeV scale SUSY particles. Values of $Z$
as small as $Z\simeq 0.01$ would give $m_{\rm DM} \simeq 100$ GeV.
Note that in this case where $\phi$ decays mainly to Higgses,
the production of $R$-parity odd particles in three-body decays
requires the heavy and/or light Higgs decay to a gaugino/Higgsino pairs to be blocked kinematically.
\een

In summary, in the branching scenario the lightest neutralino can be any mixture of Bino, Wino, and Higgsino and both thermal over and underproduction cases can be accommodated. The abundance of the QCD axion is totally negligible due to dilution by modulus decay at $T_ {\rm rh} \ll \Lambda_{\rm QCD}$.

\section{Conclusions}

In this paper we showed how sequestered LVS models give rise naturally to
non-thermal DM from the decay of the lightest modulus $\phi$.
Moreover, there is no moduli-induced gravitino problem since $m_{3/2}\simeq 10^{10}\,{\rm GeV}\gg m_\phi\simeq5\times 10^6$ GeV.
Thanks to sequestering, the superpartner spectrum is still in the TeV range even with such a heavy gravitino.
Depending on the way in which $\phi$ couples to the visible sector, there are two regimes
for the reheat temperature $T_{\rm rh}$. The case of high $T_{\rm rh}\simeq 1$ GeV
is realised when $\phi$ decays mainly to Higgses, and corresponds to the ``annihilation scenario''.
Axionic DR overproduction is avoided either by the presence of anomalous $U(1)$s which
eat dangerous axions or by allowing suitable couplings in the Giudice-Masiero term.
The resulting non-thermal DM has two components: Wino/Higgisino-like neutralinos
with masses $m_{\rm DM}> 40$ GeV and QCD axions (we note that indirect detection may limit the viability of non-thermal Wino DM \cite{winodmnonth}). 
The reheat temperature can instead be lowered to $T_{\rm rh}\simeq 10$ MeV if $\phi$ decays mainly to gauge bosons
(or if the decay to Higgses is suppressed). This is the case of the ``branching scenario''
where the QCD axion can only be an open string mode whose abundance is diluted
by the decay of $\phi$ since $T_{\rm rh} \ll \Lambda_{\rm QCD}$.
Both thermal over and underabundance cases can be accommodated in this scenario and the DM mass can vary from $\mc{O}(\rm{GeV})$ to $\mc{O}(\rm{TeV})$. %GeV

\section{Acknowledgements}

The work of B.D. is supported by DE-FG02-95ER40917. K.S. is supported by NASA Astrophysics Theory Grant NNH12ZDA001N. R.A., B.D., and K.S. would like to thank the Center for Theoretical Underground Physics and Related Areas (CETUP* 2013) in South Dakota for its hospitality and for partial support during the completion of this work.

\end{document}